\pdfoutput=1

\documentclass[aps,prc,reprint,showpacs,superscriptaddress]{revtex4-1}
\usepackage{graphicx}
\usepackage{amsmath}
\usepackage{amssymb}
\usepackage{dcolumn}
\usepackage{dsfont}
\usepackage{latexsym}
\usepackage{rotating}
\usepackage{color}
\usepackage{latexsym}
\usepackage{bbm}
\usepackage{subfigure}
\usepackage{float}
\usepackage{epsfig}
\usepackage{psfrag}
\usepackage{bm}
\usepackage{amsthm}
\usepackage{eucal}
\usepackage{mathrsfs}
\usepackage{url}
\usepackage{color}

\usepackage{hyperref}
\hypersetup{
colorlinks=true,
final=true,
linkcolor=blue,
citecolor=blue,
filecolor=blue,
urlcolor=blue,
}

\begin{document}

\title{Topological superconductivity and Majorana bound states at the LaAlO$_3$/SrTiO$_3$ interface}
\author{N. Mohanta}
\email{nmohanta@phy.iitkgp.ernet.in}
\affiliation{Department of Physics, Indian Institute of Technology Kharagpur, W.B. 721302, India}
\author{A. Taraphder}
\affiliation{Department of Physics, Indian Institute of Technology Kharagpur, W.B. 721302, India}
\affiliation{Centre for Theoretical Studies, Indian Institute of Technology Kharagpur, W.B. 721302, India}

\begin{abstract}
  The interface between two band insulators LaAlO$_3$ and
  SrTiO$_3$ exhibits low-temperature superconductivity coexisting with
  an in-plane ferromagnetic order.  We show that topological
  superconductivity hosting Majorana bound states can be induced at
  the interface by applying a magnetic field perpendicular to the
  interface.  We find that the dephasing effect of the in-plane
  magnetization on the topological superconducting state can be
  overcomed by tuning a gate-voltage.  We analyze the vortex-core
  excitations showing the zero-energy Majorana bound states and the
  effect of non-magnetic disorder on them. Finally, we propose an
  experimental geometry where such topological excitations can be
  realized.
\end{abstract}


\maketitle

\section{Introduction}
Majorana fermion, which often arises as a quasi-particle excitation in
condensed matter systems, is being studied intensively due to its
indispensable utility in defect-free topological quantum
computation~\cite{Alicea2011}.  The Majorana bound state (MBS)
naturally exists in spin-triplet chiral $p$-wave superconductivity in
superfluid He$^3$ (A-phase)~\cite{RevModPhys.69.645} and
Sr$_2$RuO$_4$~\cite{Ishida1998} and in the fractional quantum Hall
state at $5/2$ filling~\cite{Moore1991362}.  Also, there have been
several proposals of experimentally feasible systems that host MBS
such as quantum wire coupled to $s$-wave
superconductor~\cite{PhysRevLett.105.177002},
semiconductor-superconductor
heterostructure~\cite{PhysRevLett.105.077001, Mourik25052012},
proximity-induced superconductor at the surface of topological
insulator~\cite{PhysRevLett.100.096407}, 2DEG at semiconducting
quantum well~\cite{PhysRevB.81.125318}, and the more promising Al-InSb
nanowire topological superconductor~\cite{Das2012}.  Also there is a
trend to realize topological orders in fermionic $s$-wave superfluids
of ultracold atoms in optical lattices~\cite{PhysRevLett.101.160401}.
However, due to the lack of convincing experimental evidence so far,
Majorana fermion still remains as elusive and its search, therefore,
should expand on to uncharted routes, new systems and novel
experimental designs.

The two-dimensional electron liquid (2DEL) at the LaAlO$_3$/SrTiO$_3$ interface is formed as a result of an intrinsic electronic transfer mechanism known as the polar catastrophe in which half an electronic charge is transferred from the top of the polar LaAlO$_3$ to the terminating TiO$_2$ layer on the non-polar SrTiO$_3$ side to avoid a charge discontinuity at the interface~\cite{Ohtomo2004, Nakagawa2006}. The 2DEL becomes superconducting below $200$mK~\cite{Reyren31082007, GariglioJPCM2009} along with large magnetic moment ($\sim$0.3-0.4 $\mu_B$) aligned parallel to the interface plane~\cite{Li2011, Bert2011, PhysRevLett.107.056802}. Suggestions for the mechanism responsible for the novel ferromagnetism include a RKKY interaction~\cite{Michaeli_PRL2012}, a double exchange process~\cite{Banerjee2013} and Oxygen vacancies~\cite{PhysRevB.86.064431} developed at the interface during the deposition process. On the other hand, there are proposals for phonon-mediated electron-pairing~\cite{Michaeli_PRL2012,Banerjee2013,PhysRevB.89.184514} as well as unconventional superconductivity~\cite{schmalian_arxiv2014,chetan_PRB2013,caprara_arXiv2013}. Another important feature of the interface is the Rashba spin-orbit interaction (SOI) which arises because of the broken mirror symmetry along perpendicular to the interface. A back-gate voltage can tune both the electron concentration and the Rashba SOI and therefore can drive a superconductor-insulator transition~\cite{Caviglia2008,CavigliaPRL2010} making the system a potential candidate for novel electronic devices~\cite{Mannhart26032010}.

Here we show that a magnetic field, applied perpendicular to the interface plane, can induce topological superconductivity that harbours gapless edge states and MBS at the core of a vortex. The intrinsic in-plane magnetization favours a finite momentum pairing and, therefore, weakens the topological superconducting phase. We show that by tuning the Rashba SOI (\textit{i.e.} the back-gate voltage) the topological superconducting phase can be stabilized against the deterrent effect of the in-plane magnetization. We study the in-gap excitations and find that the zero-energy MBS located at the vortex-core is accompanied by low-energy particle-hole symmetric in-gap states of electronic origin. We study the effect of non-magnetic disorder on the low-energy excitations and observe that the MBS vanishes with moderate disorder. We propose an experimental set-up where the existence of MBS can be tested experimentally and discuss about some future directions.

 The large Rashba spin-orbit interaction
(RSOI), arising from the broken inversion symmetry along the
$\hat{z}$-direction, converts the $s$-wave superconductivity into an
effective \textit{p$_x$ $\pm$ ip$_y$} one. The in-plane magnetization
$h_x$ introduces asymmetry in the two-sheeted Fermi surface leading to
finite-momentum pairing of electrons~\cite{mohantaJPCM}.  The main
idea to get a topological superconductivity in two-dimensional
$s$-wave superconductor with RSOI ~\cite{PhysRevB.77.220501} is to
apply a large perpendicular Zeeman field $h_z$ to essentially remove
one of the helicities of the RSOI-induced \textit{p$_x$ $\pm$ ip$_y$}
states.  One has to circumvent the deterrent effect of the in-plane
magnetization to stabilize topological superconductivity in this
system.

In  the present  work, we  predict  that regardless  of the  asymmetry
around the  \textit{$\Gamma$}-point in  the Fermi surface,  the single
species \textit{p$_x$  $+$ ip$_y$} superconductivity  still harbours a
single MBS  at the core  of a vortex.   We show that the  tunable RSOI
competes with the in-plane  magnetization and restores the topological
phase.  However, we find  that the  excitation at  the vortex  core is
highly sensitive  to non-magnetic  disorder; even a  moderate disorder
can destroy the MBS as the magnetization breaks time-reversal symmetry
explicitly.  As the interface in LAO/STO possesses intrinsic disorders
like Oxygen vacancies, developed  during the deposition process, it is
indeed  quite challenging  to detect  a Majorana  fermion  here.  Some
remedies and possible experimental  requirements for realizing the MBS
in this system are discussed.

\section{Model for interface 2DEL}
The electrons in the 2DEL occupy the three $t_{2g}$-bands (\textit{viz.} 
d$_{xy}$, d$_{yz}$ and d$_{xz}$) of Ti atoms in the terminating TiO$_2$ 
plane giving rise to a quarter-filled ground state. Excess electrons 
supplied by the Oxygen vacancies or the back-gating accumulate on the 
next TiO$_2$ layer below the interface. Electrons in the $d_{xy}$ band 
are mostly localized at the interface sites due to Coulomb correlation. 
The electrons in the itinerant bands, in the TiO$_2$ layer just 
below the interface couple to the localized  moments via ferromagnetic 
exchange leading to an in-plane spin-ordering of the interface electrons. 
The temperature variation of the gap is found to be BCS-like,
$2\Delta_0/{k_BT_{gap}\approx3.4}$, where $\Delta_0$ is the pairing-gap 
amplitude at $T=0$, $k_{B}$ is the Boltzmann constant and $T_{gap}$ is 
the gap-closing temperature~\cite{RichterNature2013} (the 
transition to the superconducting state is of BKT-type~\cite{Caviglia2008}); 
it is therefore generally assumed that the itinerant electrons at the 
interface undergo conventional $s$-wave pairing, although there are suggestions 
of unconventional pairing as
well~\cite{schmalian_arxiv2014,chetan_PRB2013,caprara_arXiv2013}.  The
simple  model describing  the interface  electrons, at  the mean-field
level, is \vspace{-1em}
\begin{equation}\vspace{-1em}
\begin{split}
{\cal H}&=\sum_{\mathbf{k},\sigma}(\epsilon_{\mathbf{k}}-\mu) c_{\mathbf{k}\sigma}^\dagger c_{\mathbf{k}\sigma}+\alpha \sum_{\mathbf{k},\sigma,\sigma^{\prime}} [\mathbf{g_k}\cdot \mathbf{\sigma}]_{\sigma,\sigma^{\prime}} c_{\mathbf{k}\sigma}^\dagger c_{\mathbf{k}\sigma^{\prime}}\\
&-\sum_{\mathbf{k},\sigma,\sigma^{\prime}}[h_x \sigma_x]_{\sigma,\sigma^{\prime}}c_{\mathbf{k}\sigma}^\dagger c_{\mathbf{k}\sigma^{\prime}}+\sum_{\mathbf{k}} (\Delta c_{\mathbf{k}\uparrow}^\dagger c_{-\mathbf{k}\downarrow}^\dagger+h.c.)
\end{split}
\label{model}\vspace{-1em}
\end{equation}
where $\epsilon_{\mathbf{k}}$ = - 2$t$(cos $k_x$ + cos $k_y$) is the
energy band dispersion with the hopping amplitude $t$ and chemical
potential $\mu$, $\mathbf{g_k}=(\sin k_y, -\sin k_x)$ describes the
RSOI of strength $\alpha$ and
$\Delta=-<c_{\mathbf{k}\uparrow}c_{-\mathbf{k}\downarrow}>$ is the
pairing gap and $\mathbf{\sigma}=[\sigma_x, \sigma_y, \sigma_z]$ being
the Pauli matrices.  

\section{Inducing topological  superconductivity}
\begin{figure}[!ht]\vspace{-1.30em}
\begin{center}
\epsfig{file=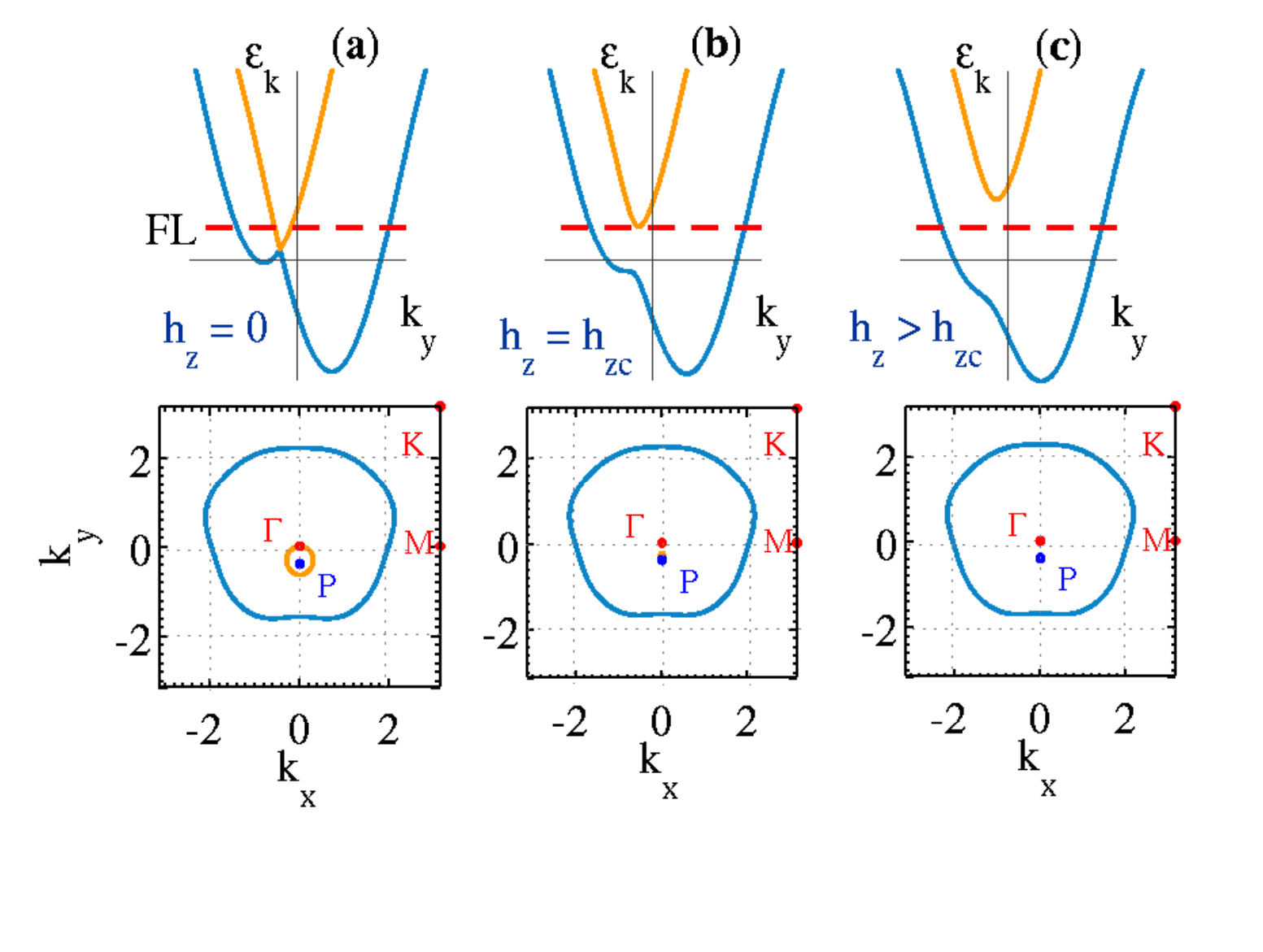,trim=0in 0.7in 0.in 0.in,clip=true, width=80mm}\vspace{-1em}
\caption{(Color online)  The band structure and Fermi surface:   
  outlining the    scheme   of    inducing    topological
  superconductivity via external Zeeman field $h_z$ with
  (a)  $h_z  =  0$:  normal  superconductivity  (b)  $h_z  =  h_{zc}$:
  transition   point    and   (c)   $h_z    >   h_{zc}$:   topological
  superconductivity.}
\label{band_fs}\vspace{-1.5em}
\end{center}
\end{figure}

The RSOI breaks the spin-degeneracy of the
original bands and creates two new electronic bands while the in-plane
magnetization $h_x$ shifts the Berry curvature from the $\Gamma$-point
to P (0, $-h_x/\alpha$) point, thus making it
energetically favorable for the electrons to pair up at finite center-of-mass 
momentum proportional to $h_x$.  In the diagonal basis of the
Rashba Hamiltonian, one essentially obtains \textit{p$_x$ $\pm$
  ip$_y$} pairing symmetry of the
superconductivity~\cite{mohantaJPCM}.  When an external Zeeman
field, perpendicular to the interface 2DEL, ${\cal
  H}_Z=-h_z\sum_{\mathbf{k}}[c_{\mathbf{k}\uparrow}^{\dagger}c_{\mathbf{k}\uparrow}
- c_{\mathbf{k}\downarrow}^{\dagger}c_{\mathbf{k}\downarrow}]$ is
applied, a gap is opened at the point P.  
The pairing amplitudes in the newly created bands, 
$\epsilon_{\pm}(\mathbf{k})=\epsilon_{\mathbf{k}}^{\prime} \pm \xi$
are given by
\vspace{-0.8em}
\begin{equation*}\vspace{-1em}
\begin{split}
&\Delta_{\pm \pm}=-\frac{\alpha}{2\xi} \Delta \left(\sin k_y \pm i\sin k_x \right): \text{intraband \textit{p}-wave}\\
&\Delta_{+-}=\frac{h_z}{\xi} \Delta: \text{interband \textit{s}-wave}
\end{split}
\end{equation*}\vspace{-0em}
where 
$\xi=\left(\alpha^2|\mathbf{g_k}|^2+h^2-2 \alpha h_x \sin k_y\right)^{1/2}$, 
$h^2=h_x^2+h_y^2$ and $\epsilon^{\prime}_{\mathbf{k}}=\epsilon_{\mathbf{k}}-\mu$. 
As shown in
FIG.~\ref{band_fs}, when $h_z$ increases beyond a critical field
$h_{zc}$, there is only one Fermi surface (\textit{i.e.}, one of the
two helicities \textit{p$_x$ $\pm$ ip$_y$} is removed) and the
superconductivity is transformed into a topological superconductivity.

\begin{figure}[!ht]\vspace{-1.2em}
\begin{center}
\epsfig{file=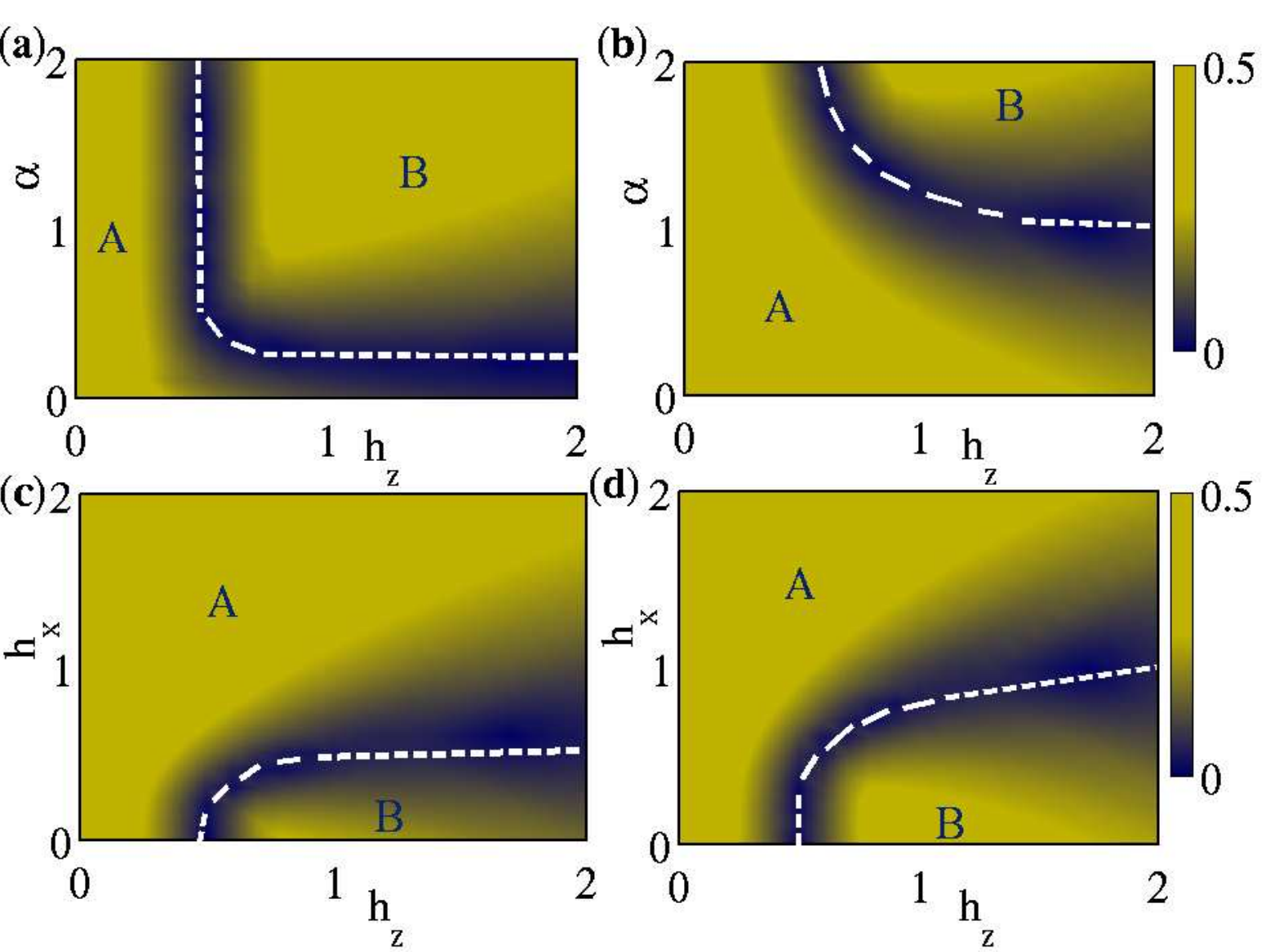, width=70mm}\vspace{-1em}
\caption{(Color online) The quasi-particle excitation gap $E_g$ in the
  $\alpha-h_z$  ((a)  and (b))  and  $h_x-h_z$  ((c)  and (d))  planes
  showing  the  quantum  phase  transition between  the  topologically
  trivial  (region  A)  and  non-trivial  (region  B)  superconductors.
  Parameters:  (a)  $h_x=0.2$, (b)  $h_x=1.0$,  (c) $\alpha=0.2$,  (d)
  $\alpha=1.0$ with $t=1$, $\mu=0$  and $\Delta=0.5$. The dashed white
  lines delineate the boundary between the two phases.}
\label{phase}\vspace{-2em}
\end{center}
\end{figure}

The effective Hamiltonian ${\cal H}_{eff}={\cal H}+{\cal H}_Z$ of the system can 
be written in the usual Nambu basis 
$\Psi(\mathbf{k})=[c_{\mathbf{k}\uparrow}, c_{\mathbf{k}\downarrow}, c_{-\mathbf{k}\downarrow}^{\dagger}, -c_{-\mathbf{k}\uparrow}^{\dagger}]$  
as
\vspace{-0.5em}
\begin{equation}\vspace{-0.5em}
\begin{pmatrix} \begin{array}{cc} {\cal H}_0(\mathbf{k}) & \Delta \\ \Delta & -\sigma_y {\cal H}_0(\mathbf{k}) \sigma_y^* \end{array} \end{pmatrix} \Psi(\mathbf{k}) = E_{\pm}(\mathbf{k}) \Psi(\mathbf{k})
\end{equation}
where ${\cal H}_0(\mathbf{k})=\epsilon^{\prime}_{\mathbf{k}}+\alpha \mathbf{g_k \cdot \sigma}-h_z\sigma_z-h_x\sigma_x$
and we obtain the bulk spectrum
$E_{\pm}^2(\mathbf{k})= \left( \epsilon^{\prime 2}_{\mathbf{k}}+\Delta^2+\xi^2 \right) \pm \zeta$
where $\zeta=\left[ \Delta^2h_z^2+\epsilon^{\prime 2}_{\mathbf{k}} \xi^2 \right]^{1/2}$. 
The transition to the topological state occurs only when the gap of the 
bulk spectrum closes \textit{i.e.}, when
$\zeta= \epsilon_{\mathbf{k}}^{\prime 2}+\Delta^2+\xi^2$.
This is satisfied (with $\Delta\not=0$) when either $\xi^2=h_z^2$ 
or $\epsilon_{\mathbf{k}}^{\prime 2}+\Delta^2=h_z^2$ which essentially reduces to the familiar 
relation $h_z  = \sqrt{\Delta^2+\mu^2}$~\cite{PhysRevB.81.125318} when $h_x=0$. 
As $h_z$ increases beyond this transition point, a  topologically  
protected excitation  gap
$E_g$  (given by  the minimum  of $E_-(\mathbf{k})$), proportional to the  
Rashba coupling strength $\alpha$ in the limit of small $h_z$, is induced. 
Therefore, $E_g$ keeps track of the quantum phase transition from ordinary 
superconductivity to topological superconductivity.
The phase diagrams of the system, revealing the parameter regime in
which the  topological superconductivity can be achieved, is shown in
FIG.~\ref{phase}. 
There are  two competing  energy gaps, the  minimum of which  tries to
destroy the topological state; one  is the induced bulk excitation gap
at       the       Fermi      level       ($\mathbf{k}=\mathbf{k_F}$),
$\Delta_{FS}=2\Delta_{--}$, the  other is the Zeeman gap  at the point
P,  given   by  $\Delta_z=E_-(0,-h_x/{\alpha})$.   In   addition,  the
in-plane  magnetization,  by  introducing a  finite-momentum  pairing,
weakens the topological superconductivity.  However, the RSOI competes
with  the in-plane  magnetization  to restore  the topological  state,
which is clearly visible in the phase diagram, FIG.~\ref{phase}. Since
the RSOI in LaAlO$_3$/SrTiO$_3$  interface is tunable by external gate
voltage~\cite{CavigliaPRL2010},  it  can  be  tuned to  stabilize  the
topological state.

\section{Vortex core excitations}
The induced topological superconductivity exhibits edge states and 
zero-energy MBS at the core of a vortex~\cite{Nagai_JPSJ2014}. 
Since Majorana fermions are essentially half an ordinary 
fermion, they always come in pair, generally located in different 
vortex-cores. A system having only one vortex, hosts the second Majorana 
fermion at the boundary. The topological property of the gapless 
excitation at the boundary is connected to the bulk topological state 
as a consequence of the ``bulk-boundary correspondence''. In the 
following we study the vortex core states by solving the following 
effective BdG Hamiltonian in real space.  
\vspace{-0.5em}
\begin{equation}\vspace{-1.5em}
\begin{split}
{\cal H}&=-t^{\prime}\hspace{-0.8em}\sum_{<ij>,\sigma}c_{i\sigma}^\dagger c_{j\sigma}-\mu\sum_{i,\sigma} c_{i\sigma}^\dagger c_{i\sigma}
-\hspace{-0.5em}\sum_{i,\sigma,\sigma^{\prime}}(\mathbf{h \cdot \sigma})_{\sigma \sigma^{\prime}}c_{i\sigma}^\dagger c_{i\sigma^{\prime}}\\
&-i\frac{\alpha}{2}\hspace{-0.5em}\sum_{<ij>,\sigma,\sigma^{\prime}}\hspace{-0.8em}(\mathbf{\sigma}_{\sigma \sigma^{\prime}} \times \mathbf{\hat{d}}_{ij})_z c_{i\sigma}^{\dagger}c_{j\sigma^{\prime}}
+  \sum_{i}(\Delta_i c_{i\uparrow}^{\dagger}c_{i\downarrow}^{\dagger}+h.c.)
\end{split}\vspace{-1em}
\label{HBdG}
\end{equation}
where $t^{\prime}$ is the hopping amplitude of electrons on a 
square lattice, $\mathbf{h}=(h_x, 0, h_z)$, $\mathbf{\hat{d}}_{ij}$ is
unit   vector   between   sites   $i$  and   $j$,   and   $\Delta_i=-U
(c_{i\uparrow}c_{i\downarrow})$ is  the onsite pairing  amplitude with
$U$, the  attractive pair-potential.  The  Hamiltonian (\ref{HBdG}) is
diagonalized via  a spin-generalized Bogoliubov-Valatin transformation
$\hat{c}_{i\sigma}(r_i)=\sum_{i,\sigma^{\prime}}u_{n\sigma\sigma^{\prime}}(r_i)\hat{\gamma}_{n\sigma^{\prime}}+v_{n\sigma\sigma^{\prime}}^*(r_i)\hat{\gamma}^{\dagger}_{n\sigma^{\prime}}$
and    the    quasi-particle    amplitudes   $u_{n\sigma}(r_i)$    and
$v_{n\sigma}(r_i)$ are determined by solving the BdG equations: ${\cal
  H}\phi_n(r_i)=\epsilon_n\phi_n(r_i)$                            where
$\phi_n=[u_{n\uparrow}(r_i),u_{n\downarrow}(r_i),v_{n\uparrow}(r_i),v_{n\downarrow}(r_i)]$.
To  model  a  vortex,  we  use  open  boundary  conditions  and  solve
self-consistently the  BdG equations by taking an initial ansatz for the
gap as  $\Delta_j=\Delta_0r_je^{i\phi_j}/\sqrt{r_j^2 + 2r_c^2}$, where
($r_j,  \phi_j$) are the polar coordinates  of site $j$
with respect to the core at the center of the 2D plane,
$\Delta_0$ and $r_c$ are respectively the depth and size of the vortex
core. The results, presented here, are for a $40\times40$ system 
with vortex-size $r_c=1$. For  the rest  of the paper, we set $\mu=0$, 
$t^{\prime}=1$ and $U=4.0$, unless explicitly specified.
\begin{figure}[!ht]\vspace{-0.7em}
\begin{center}
\epsfig{file=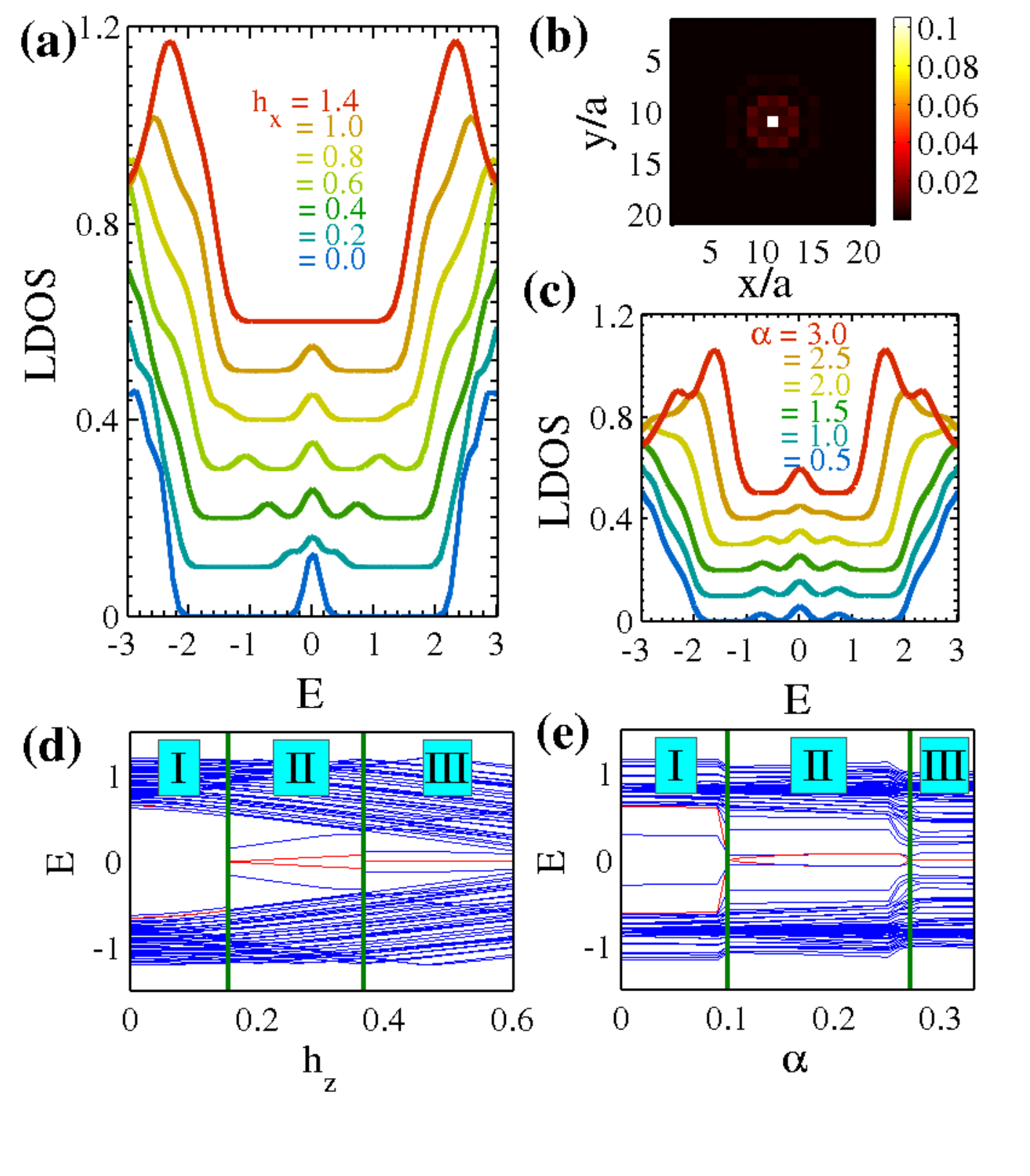, width=85mm}\vspace{-2.5em}
\caption{(Color online) (a) The LDOS at the vortex core for in-plane fields $h_x$ showing
  the MBS at zero-energy and the low-energy fermionic states for
  $\alpha=1.0$ and $h_z=1.0$.  (b) Plot  of the density of zero energy
  quasi-particles,   $|u_{n\uparrow}(r_j)|^2+|u_{n\downarrow}(r_j)|^2$,
  showing  the  vortex-core  state  on  the 2D  plane  for  $h_x=0.2$,
  $h_z=1.0$  and $\alpha=1.0$ for a $25\times25$ system.   (c)  LDOS for  various $\alpha$  with
  constant $h_x=0.2$ and $h_z=1.0$.  (d)-(e) Quasi-particle spectra as
  a  function of  $h_z$\,\,  ($\alpha=0.2$, $h_x=0.1$  and $U=1$)  and
  $\alpha$ ($h_z=0.6$, $h_x=0.2$  and $U=1$) delineating three regions
  - (I)  normal superconductivity, (II)  normal superconductivity with
  magnetized  vortex and  (III) topological  superconductivity  with a
  Majorana BS (the red lines at zero bias).}
\label{vortex}
\end{center}\vspace{-1.7em}
\end{figure}
In FIG.~\ref{vortex}(a),  we plot the local density  of states (LDOS) at the vortex core,
given                                                                by
$\rho(\epsilon,r_i)=\frac{1}{N}\sum_{n,\sigma}[|u_{n\sigma}(r_i)|^2\delta
(\epsilon-\epsilon_n)+|v_{n\sigma}(r_i)|^2\delta(\epsilon+\epsilon_n)]$
for  various in-plane  field  $h_x$,  $N$ being  the  total number  of
lattice sites.  Evidently, the zero-bias Majorana  mode is accompanied
by other low-energy vortex-bound-states which generally appear at a 
vortex core in conventional superconductors and are known as the 
Caroli-de Gennes Matricon states~\cite{Caroli_1964}. These states exist 
in both the normal and topological superconducting phase.   
With increasing  $h_x$, the fermionic states move away and mix  with 
the bulk  bands and the  MBS vanishes suddenly  as $h_x$ reaches   the   
critical  value   for   transition   to  the   trivial superconducting 
state.  The Majorana  excitation at the vortex core is shown in FIG.~\ref{vortex}(b).   
As shown in FIG.~\ref{vortex}(c), the bulk  band-gap  reduces  slowly   
with  increasing  $\alpha$  and  the low-energy  fermionic  excitations  
merge  with the  MBS.  Remarkably, similar features  of the  LDOS, for 
various  gate voltages,  have been seen  in the  tunneling  spectra  
obtained at  50  mK using  Au-gated LAO/STO tunnel device~\cite{Richter_thesis}.  
FIG.~\ref{vortex}(d)-(e) show three  regimes: regimes  I, II and  III 
are  all superconducting: region II has a magnetized  vortex and only 
region III has non-trivial superconductivity and the Majorana modes 
(two red lines at zero energy for the vortex-core and edge  excitations). 
Region II is quite interesting as it has a  zero energy fermionic excitation 
close to  the boundary of I which  arises   in  this   trivial  
superconducting  state   from  the competition between  the Zeeman field  
that magnetizes the  vortex and the superconductivity, thereby changing 
the vortex structure~\cite{SchafferPRB2013}. The spacing between the in-gap bound states is typically of the order of $\Delta^2/E_F$, where $\Delta$ is the gap amplitude, $E_F$ is the Fermi energy. In the present situation, the critical temperature of the real system is not proportional to the mean-field 'gap' we are using~\cite{RichterNature2013}: it is dictated primarily by the coupling between the superconducting grains and therefor much smaller (200 mK) than would otherwise appear from the gap values. In the tunneling spectrum, the position of these 
in-gap bound states is thus not completely determined by the real T$_c$ (200mK). However, while tuning the perpendicular magnetic field ($h_z$) or 
the gate-voltage (\textit{i.e.}, Rashba SOI strength $\alpha$), the abrupt 
transition from region II to region III, in FIG.3(d)-(e), can also give rise to 
a stronger, unique experimental signature of Majorana bound states.
 
\section{Influence  of   disorder}
\begin{figure}[!ht]\vspace{-1em}
\begin{center}
\epsfig{file=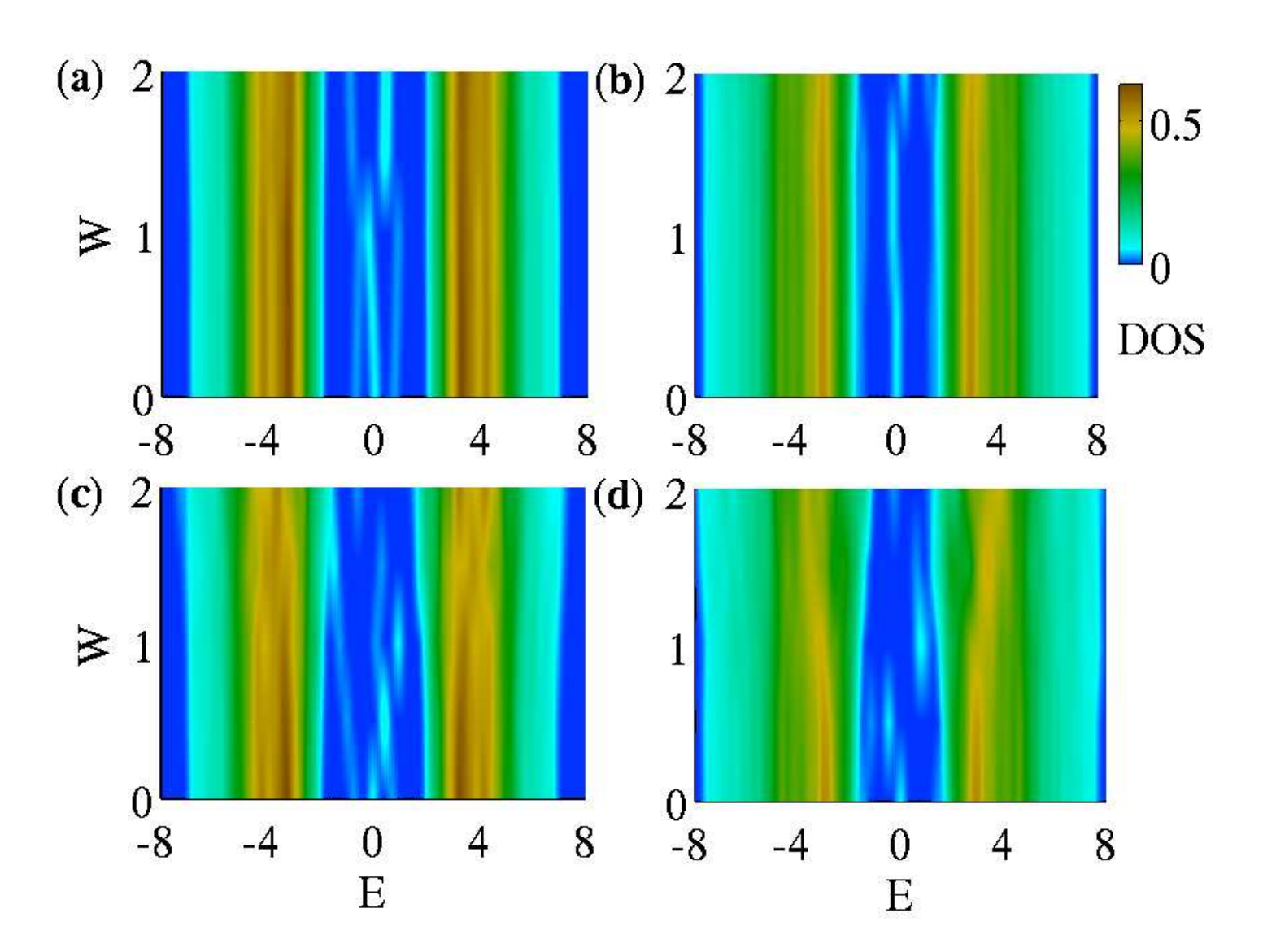, width=80mm}\vspace{-1em}
\caption{(Color  online) Plot  of the  density of states  as a  function of
  disorder  strength $W$,  revealing the  behaviour of  the low-energy
  excitations inside  the bulk  superconducting gap (a)  $h_z=0.5$, $h_x=0.4$,
  (b) $h_z=0.5$, $h_x=1.0$, (c) $h_z=1.0$, $h_x=0.4$ and (d) $h_z=1.0$, $h_x=1.0$. 
Other parameters are same as in FIG.~\ref{vortex}.}
\label{dis}
\end{center}\vspace{-1em}
\end{figure}

The topological excitations are,
however, quite fragile against the imperfections of the host system. 
The LaAlO$_3$/SrTiO$_3$ interface contains intrinsic  disorder such as 
Oxygen vacancies, known to have significant effects on the interface
electrons~\cite{PhysRevB.86.064431,  MohantaVacancyJPCM2014}. It is,
therefore, imperative to study the robustness of the MBS against
non-magnetic disorder, introduced through an onsite random potential $V_d$ in 
the Hamiltonian~(\ref{HBdG}) by ${\cal H}_{dis}=V_d \sum_{i,\sigma}
c_{i\sigma}^\dagger  c_{i\sigma}$, where, $V_d \in [-W, W]$ uniformly. In 
FIG.~\ref{dis}, we plot the density of states
$\rho(\epsilon)=\frac{1}{N}\sum_{n,r_i,\sigma}[|u_{n\sigma}(r_i)|^2\delta
(\epsilon-\epsilon_n)+|v_{n\sigma}(r_i)|^2\delta(\epsilon+\epsilon_n)]$
for disorder realizations of 
various disorder strength $W$. The MBS is quickly destroyed 
as disorder increases and other in-gap excitations
appear within the bulk gap due to the defects. The MBS is, in fact,
not expected to be robust against perturbations like disorder in this
system since the time-reversal symmetry is already broken explicitly 
~\cite{PhysRevB.83.184520}.
\begin{figure}[!ht]\vspace{-1em}
\begin{center}
\epsfig{file=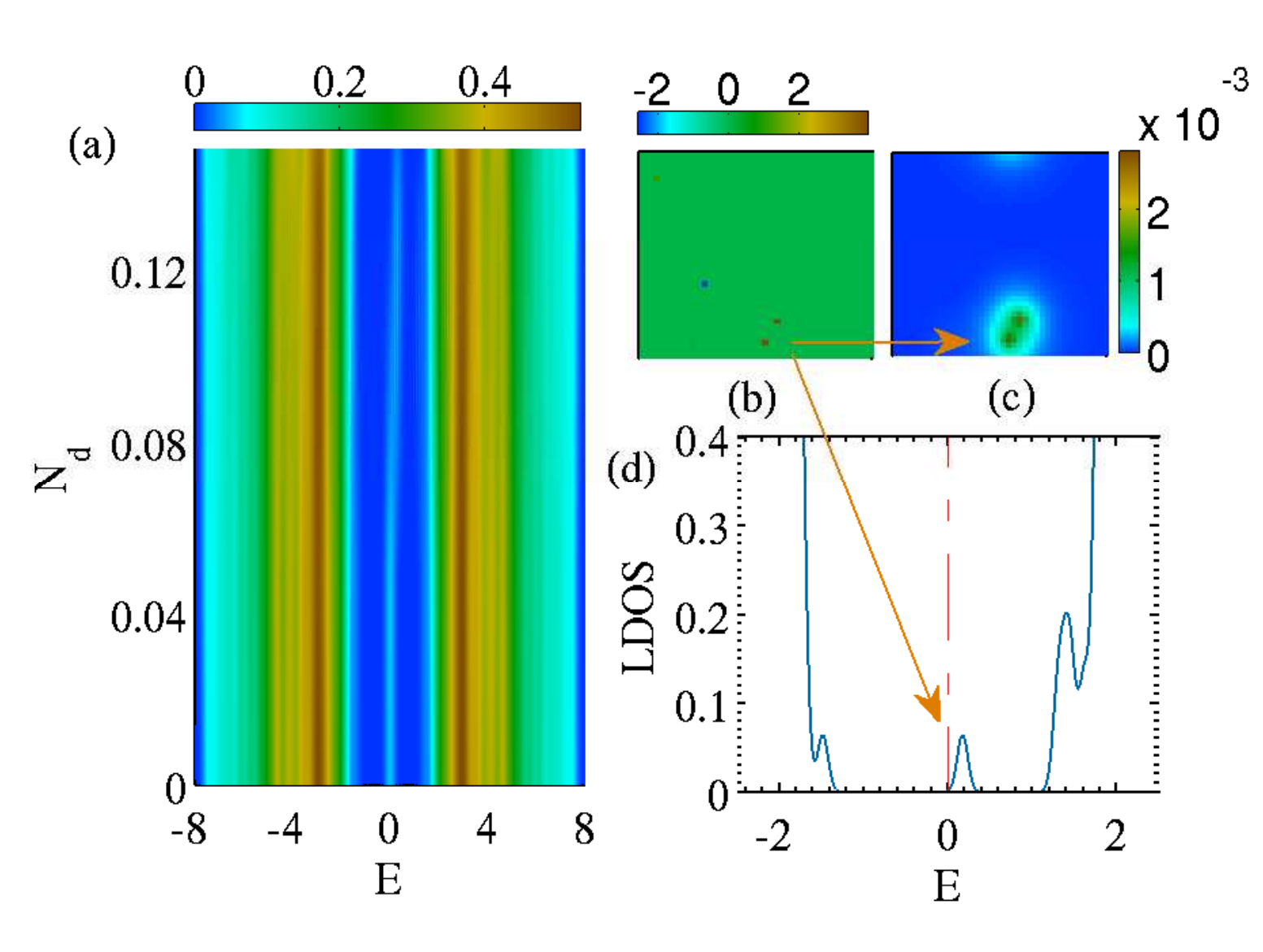, width=80mm}\vspace{-1em}
\caption{(Color  online) (a) Variation of the density of states with
disorder concentration $N_d$ with $W=1$, $h_z=0.5$, $h_x=1.0$. (b)
Disorder configuration in the 2D plane with $h_z=0.5$, $h_x=1.0$, $N_d=0.003$ and $W=2$.
(c) and (d) are respectively the density of quasi-particle $i.e.$
$|u_{n\uparrow}(r_j)|^2+|u_{n\downarrow}(r_j)|^2$ and the LDOS at the
defect site indicated by the arrow. Other parameters are same as in
FIG.~\ref{vortex}.}
\label{dis1}
\end{center}\vspace{-1em}
\end{figure}
As shown in FIG.~\ref{dis}, the vulnerability of the zero-energy MBS is
worse in presence of larger magnetic fields. In other words, the MBS 
survives against larger strength of disorder when $h_z$ is smaller (provided,
$h_z$ should always be greater than the critical value $h_{zc}$ to ensure
topological regime).
In the present  system, the
degree of  vulnerability is severe  due to the  in-plane magnetization
which, in reality, weakens the topological state. Hence the low-energy
excitations are destroyed even  when there is a finite superconducting
gap.
In FIG.~\ref{dis}, we show a situation where disorder of random strength (up to $W$) is present at all sites. We also consider a diluted situation by putting disorder at some random sites. FIG.~\ref{dis1}(a) describes how the low-energy in-gap excitations are affected as the disorder concentration ($N_d$) is 
varied. We find that the results are not different qualitatively from that 
in FIG.~\ref{dis}. The zero-energy Majorana excitation remains unaffected unless a 
defect potential appears exactly at the vortex-core and, as in FIG.~\ref{dis}, new 
states appear within the bulk-gap. To check where these defect-induced states are 
localized, we plot, in FIG.~\ref{dis1}, the quasi-particle density and LDOS 
spectra at a defect site. We find that, as reported previously ~\cite{Nagai_arXiv2014}, the new 
in-gap states are actually located at the defect sites. Therefore in the 
tunneling conductance measurement, getting an excitation at zero-energy does 
not necessarily confirm a Majorana particle. One has to be very careful in 
disentangling the Majorana fermion from defect induced states or Andreev 
bound states~\cite{Stanescu_PRB2013}.

\section{Experimental  aspects}
\begin{figure}[!ht]\vspace{-0.5em}
\begin{center}
\epsfig{file=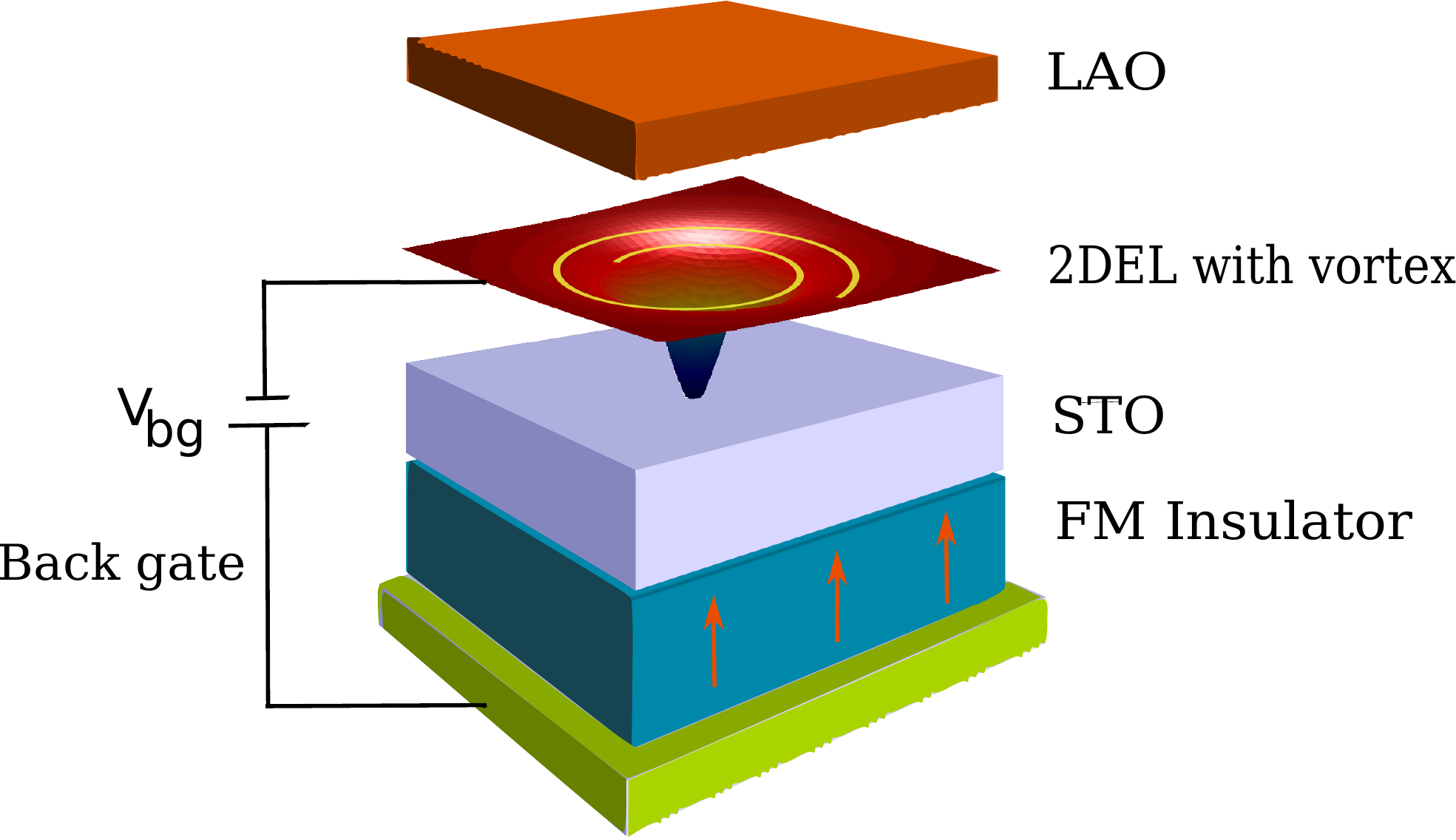, width=70mm}\vspace{-1em}
\caption{(Color online)  Schematic picture  of the proposed  setup for
  realization of MBS in LaAlO$_3$/SrTiO$_3$ interface.}
\label{design}
\end{center}\vspace{-2em}
\end{figure}

For  the  experimental realization  of
topological  superconductivity  and  MBS  in the  interface  2DEL,  an
external  Zeeman field  $h_z$ is  required.  This  can be  achieved by
mounting the  LaAlO$_3$/SrTiO$_3$ interface on top  of a ferromagnetic
insulator whose spins are aligned to the $\hat{z}$-direction, as shown
in FIG.~\ref{design}. The two  major obstacles towards a realization of
the  topological  state  however,   are  (i)  the  intrinsic  in-plane
magnetization and (ii) the intrinsic defects.
To  stabilize  the  topological  state,  the effect  of  the  in-plane
magnetization can be partially offset  by tuning the RSOI via the gate
voltage. To reduce the laser-induced defects (in pulsed laser deposition method) and intrinsic disorder such
as the  Oxygen vacancies, molecular  beam epitaxy techniques  are used
instead. Ozone  may be  used as oxidant  instead of Oxygen  during the
annealing   process   as   suggested   by   Warusawithana   \textit{et
  al.}~\cite{Warusawithana2013}.  High  resolution scanning  tunneling
microscopy (STM)  or point-contact spectroscopy may  identify a single
MBS  at  the  vortex  core.  Typical experimental  resolution  in  the
tunneling  experiments is about  2$\mu$V and  should be  sufficient to
identify  the  zero-bias peak  of  the  Majorana  BS from  the  nearby
excitations.  Though  the  mean-field  gap  is known  to  be  a  gross
over-estimation,  a  rescaling  of  the  gap in  FIG.~\ref{vortex}(c)  on  to  the
experimentally  observed gap, about  80~$\mu$V, gives  a value  of the
resolution-limit about  13~$\mu$V, close to the experimental
values ($\sim 9.5\mu$V) observed~\cite{Richter_thesis}. It is worth mentioning 
that the usual temperature range, in which the thermal fluctuation is 
small for the detection of the MBS, is less than 100 mK~\cite{Mourik25052012, Das2012} which is far 
below the Curie temperature (200 mK) of this interface superconductivity.

Recently,  it has  been shown  that superconductivity  is  possible in
quasi-1D     structures,    grown    at     the    LaAlO$_3$/SrTiO$_3$
interface~\cite{2010arXiv1009.2424C,   2012arXiv1210.3606V}   and  may
support   Majorana  zero   modes   at  the   ends   of  such   quantum
wires~\cite{PhysRevB.84.195436}.   These   are   the   steps   towards
developing qubits,  using this interface, which  is to be  used as the
building blocks of a topological quantum simulator.

\section{Conclusion}
In  summary,  we   have  shown  that  topological
superconductivity  hosting Majorana  fermions can  be realized  in the
two-dimensional  metallic interface of  LaAlO$_3$ and  SrTiO$_3$ under
the  right conditions.  The  phase diagrams  show  that an  additional
Zeeman  field  with  large  RSOI  is  required  to  achieve  a  stable
topological superconductivity.  However, non-magnetic disorder such as
Oxygen vacancies  are detrimental  to the topological  excitations. An
experimental  design   likely  to  produce   conditions  conducive  to
observing Majorana excitations is proposed.

\section{Acknowledgements} 
The authors  thank C.  Richter for sending
his unpublished  data and  noticing the similarity  with FIG. 3.  NM
thanks  J.  Mannhart  for  useful discussions. 


\end{document}